\documentstyle[11pt,newpasp,twoside,epsf]{article}
\def\edcomment#1{\iffalse\marginpar{\raggedright\sl#1\/}\else\relax\fi}
\reversemarginpar

\begin{document}
\title{Subparsec-scale HI in the nucleus of NGC~4151}
 \author{C.G. Mundell$^1$, J.M. Wrobel$^2$, A. Pedlar$^3$, J.F. Gallimore$^4$}
\affil{$^1$ARI, Liverpool John Moores University, U.K. (cgm@astro.livjm.ac.uk)}
\affil{$^2$NRAO, P.O. Box O, Socorro, NM 87801}
\affil{$^3$JBO, Macclesfield, Cheshire, U.K.}
\affil{$^4$NRAO, 520 Edgemont Rd, Charlottesville, VA 22903}

\begin{abstract}
We present sensitive, high-resolution $\lambda$21-cm VLBA+VLA
observations of the radio jet and nuclear HI absorption in NGC~4151.
The 25-mas (1.6-pc) resolution continuum image reveals a highly
collimated radio jet, underlying the discrete components seen
previously with MERLIN and the VLA.  Spatially and kinematically
complex HI absorption is detected against the whole 3-pc extent of the
continuum component predicted by Ulvestad et al. to contain the
AGN. Instead, we suggest the component against which the absorption is
detected is part of the eastern counterjet, ruling it out as the
location for the AGN.

\end{abstract}

\section{Introduction}
$\lambda$21-cm MERLIN observations of the Seyfert galaxy NGC~4151
revealed localized and marginally resolved HI absorption, with a peak
column density of $N_{\rm H}$~$\sim$6$\times$10$^{19}$~$T_{\rm S}$
cm$^{-2}$, against the component (C4) in the radio jet which is
thought to contain the AGN (Mundell et al. 1995); no absorption was
detected against the other jet components (Fig. 1a).  An east-west
column density gradient was observed and, in combination with UV
column densities and early VLBI images (Fig. 1b), which showed C4 to
consist of two components (C4E and C4W), led Mundell et al. (1995) to
suggest that the weaker, western component, C4W, contains the
optical/UV nucleus and the HI absorption is taking place against the
first component of the counterjet (C4E), due to gas in the obscuring
torus (Fig.  1c). Structural and spectral index information obtained
from subsequent radio continuum VLBA observations of the jet at 1.6
and 5~GHz led Ulvestad et al. (1998) to suggest a similar model but
with the AGN located at the emission peak of C4E; this model predicted
an absence of HI absorption against the AGN, with HI absorption
occurring only against the start of the counterjet, i.e.  the tail of
emission extending eastwards of the peak in C4E.

\section{Preliminary Results}

As shown in Fig. 1(f), HI absorption is detected against the whole
extent of C4E, thereby ruling it out as the location of the AGN. The
absorption is spatially and kinematically complex with column
densities $N_{\rm H}$~$\sim$10$^{20}$$T_{\rm S}$, where $T_{\rm S}$ is
the spin temperature for which the value is unknown but is typically
$\sim$10$^2$$-$10$^4$ K in the Galaxy (Heiles \& Kulkarni, 1988).  The
VLBA+VLA continuum image also reveals a highly collimated radio jet,
imaged over its full extent of $\sim$2\farcs5 at milliarcsecond
resolution (Fig. 1d,e), which, although less intrinsically luminous,
seems to resemble those in quasars and radio galaxies; the brighter
knots might correspond to shock features, coinciding with changes in
direction, due to interaction with the ISM in the central $\sim$100
pc.

\begin{figure}  
\plotfiddle{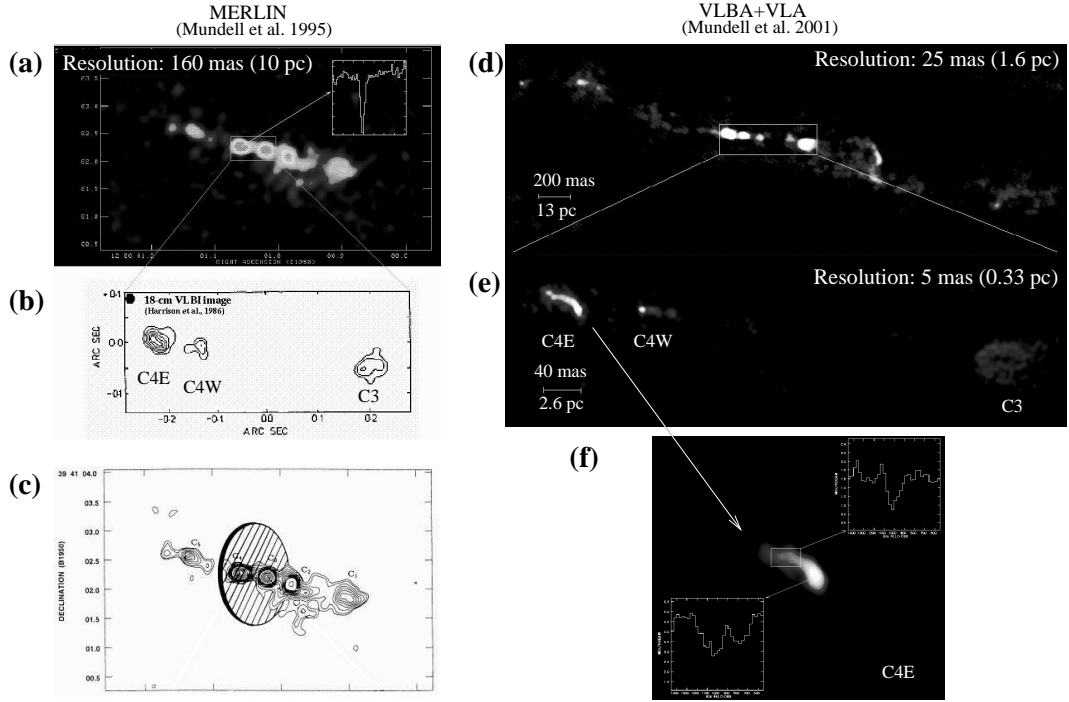}{8.2cm}{270}{77}{77}{-305}{358}
\caption{\small {\bf MERLIN:} (a) $\lambda$21-cm continuum image of radio
jet with HI absorption against component C4 inset. (b) VLBI image of
Harrison et al. with C4 resolved into C4E and C4W. (c) Model to
explain HI absorption against first component in eastern counterjet
(Mundell et al. 1995); {\bf VLA+VLBA:} (d) New $\lambda$21-cm continuum
image of jet. (e) Full resolution image of components C4 and C3
showing same region as in (b). (f) HI absorption detected against full
extent of C4E.}
\end{figure}

\end{document}